# Spatial differentiation and integration of optical beams in a slab waveguide by a single dielectric ridge

Evgeni A. Bezus, Leonid L. Doskolovich, Dmitry A. Bykov, and Victor A. Soifer

*Abstract*—We show that a very simple structure consisting of a single subwavelength dielectric ridge on the surface of a slab waveguide enables spatial differentiation and integration of the profile of optical beams propagating in the waveguide. The differentiation and integration operations are performed in transmission and in reflection, respectively, at oblique incidence of the beam impinging on the ridge. The implementation of these operations is associated with the resonant excitation of a cross-polarized eigenmode of the ridge. At the considered parameters of the structure and incidence geometry, neither "parasitic" out-of-plane scattering nor polarization conversion occurs. The presented rigorous numerical simulation results confirm high-quality differentiation and integration. We demonstrate that by choosing the quality factor of the resonance, one can achieve the required tradeoff between the differentiation (or integration) quality and the amplitude of the resulting beam. The proposed integrated structure may find application in ultrafast all-optical analog computing and signal processing systems.

*Index Terms*—Optical computing, integrated optics, optical waveguides, optical diffraction, resonance light scattering.

## I. Introduction

OVER the past few years, analog optical computing has attracted increasing attention, since it offers high-performance solution of several important computational tasks. Among the basic operations of analog processing of optical signals are spatial differentiation and integration of the profile of an optical beam. Traditionally, for the implementation of these operations, bulky optical systems consisting of lenses and filters are utilized [1]. Recently, spatial differentiators and integrators based on nanophotonic structures having a thickness comparable to the wavelength of the processed optical signal were proposed. In particular, in [2]–[5], compact analogues of the optical Fourier correlator are considered, in which the spatial filter comprises a metasurface encoding the complex transmission coefficient of a differentiating or an integrating filter. In [2], [6]–[12], various resonant structures containing systems of homogeneous layers [2], [6]–[10] and diffraction gratings [11], [12] are used for spatial differentiation and integration of optical beams. The possibility to utilize such structures for optical differentiation or integration is based on the fact that the Fano profile describing the reflection or the transmission coefficient of the structure in the vicinity of the resonance can approximate the transfer function of a differentiating or an integrating filter.

Development of planar (on-chip) differentiators and integrators, in which the processed optical signal propagates in some guiding structure, is of great interest. In particular, in a wide class of planar (integrated) optoelectronic systems, spectral or spatial filtering of optical signals is performed in a slab waveguide ("insulator-on-insulator" platform) [13], [14]. In this case, the processed signal corresponds to a superposition of slab waveguide modes with different propagation directions (in the case of spatial filtering) or with different frequencies (in the case of spectral filtering). In a recent work, the present authors proposed a simple planar differentiator consisting of two grooves on the surface of a slab waveguide and operating in reflection [15]. The differentiator operation is associated with the excitation of an eigenmode localized between the grooves. A similar graphene-based spatial integrator operating in transmission was considered in [10].

In this work, we for the first time show that a very simple structure consisting of a single subwavelength ridge on the surface of a slab waveguide can be used as a spatial differentiator and integrator. The implementation of these operations is associated with the resonant excitation of a cross-polarized eigenmode of the ridge. The presented results of rigorous numerical simulations confirm high accuracy of spatial differentiation and integration.

## II. Diffraction of slab waveguide modes on a ridge

In order to explain the resonant effects that make it possible to implement spatial differentiation and integration, let us first discuss the diffraction of slab waveguide modes on the ridge located on the surface of a slab waveguide. The geometry of

Manuscript received March 30, 2018. This work was supported by Russian Foundation for Basic Research under Grant 16-29-11683 (numerical investigation of the optical properties of the ridge, Section II), Federal Agency of Scientific Organizations under Agreement 007-GZ/Ch3363/26 (theoretical analysis of the resonances of the ridge using the effective index method, Section II), and by Russian Science Foundation under Grant 14-19-00796 (application of the structure to spatial differentiation and integration of optical beams, Sections III–V).

The authors are with the Image Processing Systems Institute of RAS — Branch of the FSRC "Crystallography and Photonics" RAS, 151 Molodogvardeyskaya st., Samara 443001, Russia, and with the Samara National Research University, 34 Moskovskoe shosse, Samara 443086, Russia (e-mail: evgeni.bezus@gmail.com).

the structure is shown in Fig. 1. For the analysis, the following parameters were chosen: refractive index of the waveguide core layer and of the ridge $n_c = 3.3212$ (GaP at the wavelength $\lambda = 630\,\text{nm}$), refractive indices of the substrate and superstrate $n_{sub} = 1.45$ (fused silica) and $n_{sup} = 1$, respectively, waveguide thickness $h_c = 80\,\text{nm}$, waveguide thickness at the ridge $h_r = 110\,\text{nm}$. In this case, at $\lambda = 630\,\text{nm}$ the waveguide is single-mode for both TE- and TM-polarizations, and the effective refractive indices of the modes amount to $n_{wg,TE} = 2.5913$ and $n_{wg,TM} = 1.6327$, respectively. Effective refractive indices of the modes at $h_r = 110\,\text{nm}$ equal $n_{r,TE} = 2.8192$ and $n_{r,TM} = 2.1867$.

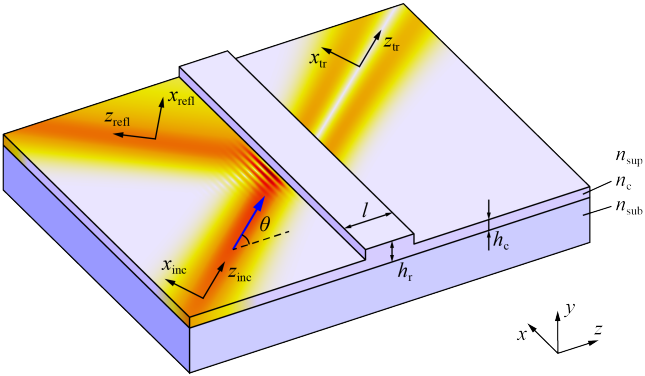

Fig. 1. Geometry of the problem of diffraction of a waveguide mode on a dielectric step.

Fig. 2 shows the reflectance $|R_{TE}(\theta,l)|^2$ and the transmittance $|T_{TE}(\theta,l)|^2$ of the ridge vs. the angle of incidence $\theta$ and the ridge length $l$ in the case of an incident TE-polarized mode, where $R_{TE}(\theta,l)$ and $T_{TE}(\theta,l)$ are the complex reflection and transmission coefficients, respectively. The plots were calculated using an in-house implementation of the aperiodic rigorous coupled-wave analysis (aRCWA) technique [16], [17]. The RCWA, also called the Fourier modal method, is an established numerical technique for solving Maxwell's equations.

It is evident that several resonances (sharp reflectance maxima and transmittance minima) are present in the spectra of Fig. 2. The resonant reflectance peaks (transmittance dips) are located in the angle of incidence range $39.05° < \theta < 57.55°$, which is marked with dashed lines in Fig. 2. It is important to note that the transmittance strictly vanishes at the resonances, and therefore the reflectance reaches unity.

Let us explain these effects. In a general case, at oblique incidence of a TE-polarized guided mode on a ridge located on the waveguide surface, reflected and transmitted TE- and TM-polarized modes are generated, as well as a continuum of plane waves arising from "parasitic" scattering out of the guiding layer to the superstrate and substrate. However, if the angle of incidence exceeds a certain value $\theta_{cr,1}$, the reflected and transmitted fields contain only the TE-polarized mode, and no out-of-plane scattering and polarization conversion occur [15], [18]. Indeed, let us denote by $k_{x,inc} = k_0 n_{wg,TE} \sin\theta$ the x-component of the wavevector of the incident TE-mode, which is parallel to the ridge boundaries. Here, $k_0 = 2\pi/\lambda$ is the wavenumber. According to the boundary conditions for the Maxwell's equations, the $k_{x,inc}$ component has to be conserved, i.e. it is the same for all the waves constituting the reflected and the transmitted fields, including the radiation scattered out of the waveguide. Therefore, at angles of incidence $\theta > \theta_{cr,1} = \arcsin(n_{wg,TM}/n_{wg,TE}) = 39.05°$, reflected and transmitted TM-polarized modes become evanescent. At $\theta > \theta_{cr,1}$, $k_{x,inc}$ also exceeds the wave vector magnitudes of the propagating plane waves over and under the waveguide (in the regions with the refractive indices $n_{sup}$ and $n_{sub}$, respectively) and therefore the incident TE-mode is not scattered out of the waveguide layer. Thus, at $\theta > \theta_{cr,1}$ the reflection and transmission coefficients corresponding to the TE-polarized mode satisfy the equality $|R_{TE}|^2 + |T_{TE}|^2 = 1$ according to the energy conservation law, since the materials of the structure are assumed to be lossless. The angle $\theta_{cr,1}$ is shown with a dashed line in Fig. 2, which corresponds to the upper boundary of the "resonance region". The lower boundary of this region corresponds to the cutoff angle of the TM-polarized mode in the ridge region, which amounts to $\theta_{cr,2} = \arcsin(n_{r,TM}/n_{r,TE}) = 57.55°$.

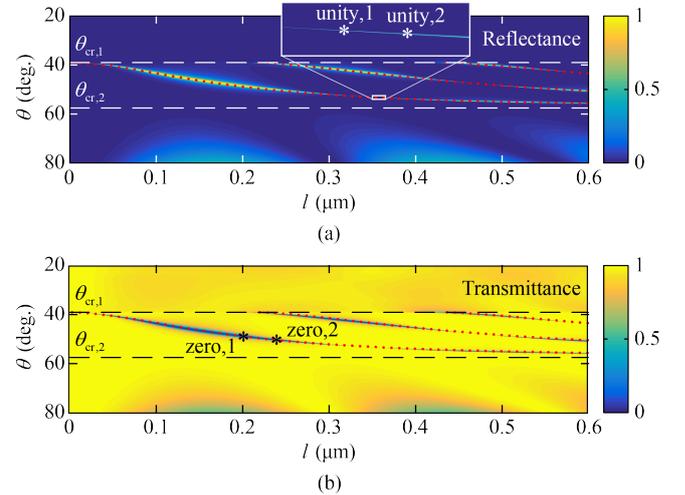

Fig. 2. Reflectance (a) and transmittance (b) of the ridge vs. the ridge length $l$ (horizontal axis) and the angle of incidence $\theta$ (vertical axis). Dashed lines show the boundaries of the resonance region and correspond to the cutoff angles of the TM-modes outside the ridge (at $h_c = 80\,\text{nm}$, upper line) and inside the ridge ($h_r = 110\,\text{nm}$, lower line). Dotted curves show the dispersion of the cross-polarized modes of the ridge. The points marked with white and black asterisks correspond to the spatial differentiation and integration examples considered below, respectively.

The presented analysis suggests that the resonances in Fig. 2 are associated with the excitation of TM-like eigenmodes of the ridge, which in this case acts as a leaky rib waveguide. Within the framework of the effective index method (EIM) [19], the TM-like modes of a rib waveguide can be approximately described by the dispersion relation of TE-polarized modes of a dielectric slab waveguide with the thickness $l$, in which the values $n_{r,TM}$ and $n_{wg,TM}$ are used as the refractive indices of the core layer and claddings, respectively. The dispersion relation of a symmetric slab waveguide can be written in the following form [20]:

$$l = \frac{\pi m + \arg r(\theta)}{k_0 n_{r,TM} \cos \theta}, \quad (1)$$

where the integer $m$ is the mode order, $r(\theta)$ is the complex reflection coefficient of a TE-polarized plane wave from the interface between the media with refractive indices $n_{r,TM}$ and $n_{wg,TM}$. The EIM-based dispersion relation (1) can be made more accurate by using the reflection coefficient of the TM-polarized mode $r_{mode,TM}(\theta)$ from the interface between two waveguides with the thicknesses $h_r = 110$ nm and $h_c = 80$ nm instead of the plane-wave reflection coefficient $r(\theta)$. The dispersion curves $l = l(\theta)$ corresponding to this refinement are shown in Fig. 2 with dotted curves and are in an excellent agreement with the resonance locations. This confirms the hypothesis that the resonant features in the spectra of Fig. 2 are indeed due to the excitation of the cross-polarized modes of the ridge.

Since in the resonant regime (at $\theta_{cr,1} < \theta < \theta_{cr,2}$) there is no out-of-plane scattering and polarization conversion in the reflected and transmitted radiation, the transmission coefficient strictly vanishes at the resonances [21], [22]. Moreover, it is evident from Fig. 2 that at different ridge lengths the resonant peaks (dips) have different angular widths (different quality factor), which vary from units to thousandths of a degree and less. This change in the quality factor is caused by the interaction between the resonances of two types: Fabry-Perot resonance of the TE-polarized mode and the guided-mode resonance of the TM-polarized mode. When two types of resonances (two modes) interact, the so-called matrix Fabry-Perot resonances occur, which can have a very high Q-factor [23]. In fact, the Q-factor in the considered structure can reach infinity, i.e. the structure supports the so-called bound states in the continuum (BICs) [24]. A detailed theoretical investigation of the Q-factor of the resonances and of the BICs in the studied structure will be the subject of a separate publication.

### III. TRANSFORMATION OF THE PROFILE OF A BEAM PROPAGATING IN THE WAVEGUIDE UPON DIFFRACTION ON A RIDGE

Let us now consider the transformation of the spatial profile of a TE-polarized beam propagating in the waveguide, which occurs upon reflection from the ridge and propagation through the ridge. In the coordinate system associated with the incident beam $(x_{inc}, z_{inc})$ (Fig. 1), the obliquely incident beam with the angle of incidence $\theta_0$ can be represented as a superposition of slab waveguide modes with different spatial frequencies $k_{x,inc} = k_0 n_{wg,TE} \sin \theta_{inc}$, where $\theta_{inc}$ is the angle between the propagation direction of the mode and the $z_{inc}$ axis. We assume that the spatial spectrum of the incident beam $G(k_{x,inc}), |k_{x,inc}| \leq g$ is narrow enough so that $g \ll k_0 n_{wg,TE}$. In this case, the field of the beam can be represented as [15]

$$u_{inc,wg}(x_{inc}, z_{inc}) = \exp(ik_0 n_{wg,TE} z_{inc}) P_{inc,wg}(x_{inc}) = $$
$$= \exp(ik_0 n_{wg,TE} z_{inc}) \int G(k_{x,inc}) \exp(ik_{x,inc} x_{inc}) dk_{x,inc}, \quad (2)$$

where $P_{inc,wg}(x_{inc})$ is the profile of the beam in a certain plane inside the waveguide, e.g. at $y = h_c/2$. The transformation of the profile of the beam upon diffraction on a ridge can be described in terms of the linear system theory. The transfer functions (TFs) describing the transformation of the beam profile in reflection and transmission have the form [6], [15]

$$H_R(k_{x,inc}) = R_{TE}(k_x(k_{x,inc})),$$
$$H_T(k_{x,inc}) = T_{TE}(k_x(k_{x,inc})), \quad (3)$$

where $R_{TE}(k_x)$ and $T_{TE}(k_x)$ are the complex reflection and transmission coefficients of the ridge, respectively, and $k_x = k_x(k_{x,inc})$ are the spatial frequencies of the TE-polarized modes constituting the incident beam written in the coordinate system associated with the structure:

$$k_x(k_{x,inc}) = k_0 n_{wg,TE} \sin(\theta_{inc} + \theta_0) \approx k_{x,inc} \cos \theta_0 + k_{x,0}, \quad (4)$$

where $k_{x,0} = k_0 n_{wg,TE} \sin \theta_0$. According to (3) and (4), the profiles of the reflected and transmitted beams in the respective coordinate systems (Fig. 1) have the form

$$P_{refl}(x_{refl}) = \int G(k_{x,inc}) R(k_x(k_{x,inc})) \exp(ik_{x,inc} x_{refl}) dk_{x,inc},$$
$$P_{tr}(x_{tr}) = \int G(k_{x,inc}) T(k_x(k_{x,inc})) \exp(ik_{x,inc} x_{tr}) dk_{x,inc}. \quad (5)$$

### IV. SPATIAL DIFFERENTIATION OF A BEAM PROPAGATING IN THE WAVEGUIDE

Let us discuss the application of the considered structure for the computation of the spatial derivative of the incident beam profile. As shown in Section II, the transmission coefficient of the ridge strictly vanishes at the resonances, which enables using this structure as an optical differentiator operating in transmission [6], [15]. Indeed, let us assume that the transmission coefficient $T_{TE}(k_x)$ vanishes at certain angle of incidence $\theta_0$ and ridge length $l$. Expanding the TF $H_T(k_{x,inc})$ of Eq. (3) into Taylor series at $k_{x,inc} = 0$ up to the linear term, we obtain:

$$H_T(k_{x,inc}) \approx \alpha_T k_{x,inc}. \quad (6)$$

Thus, in the first approximation this TF is proportional to the TF of an exact differentiator $H_{diff}(k_{x,inc}) = ik_{x,inc}$.

As an example, Fig. 3(a) shows the amplitude (absolute value) and phase (argument) of the TF $H_T(k_{x,\text{inc}})$ calculated at the point $(l_{\text{zero},1},\theta_{\text{zero},1}) = (0.2\ \mu\text{m}, 48.62°)$ marked with a black asterisk in Fig. 2(b), where $l_{\text{zero},1}$ and $\theta_{\text{zero},1}$ are the ridge length and the angle of incidence, at which the transmission coefficient vanishes. Fig. 3(a) demonstrates that in the vicinity of the point $k_{x,\text{inc}} = 0$ (i.e. at $\theta_0 = \theta_{\text{zero},1}$) the TF of the ridge is in good agreement with the TF of an exact differentiator. Let us note that the linear phase of the TF leads only to the shift of the transmitted beam similarly to the Goos–Hänchen effect and does not affect the differentiation quality.

Consider the differentiation of a beam propagating in the waveguide and having a Gaussian profile $P_{\text{inc}}(x_{\text{inc}}) = \exp(-x_{\text{inc}}^2/\sigma^2)$. The normalized spectrum of the beam $G(k_{x,\text{inc}}) \sim \exp(-k_{x,\text{inc}}^2\sigma^2/4)$ at $\sigma = 25\ \mu\text{m}$ is shown with a dashed curve in Fig. 3(a). The $1/e^2$ widths of the beam $P_{\text{inc}}(x_{\text{inc}})$ and of its spectrum $G(k_{x,\text{inc}})$ amount to $2\sqrt{2}\sigma = 70.7\ \mu\text{m}$ and $2g = 4\sqrt{2}/\sigma = 0.23\ \mu\text{m}^{-1}$, respectively. In order to assess the differentiation quality, we calculated the profile of the transmitted beam $P_{\text{tr}}(x_{\text{tr}})$ using (5). Fig. 3(b) shows the incident beam (dotted curve), the absolute value of the calculated profile of the transmitted beam (solid curve), and the absolute value of the analytically calculated derivative of the Gaussian function (dashed curve). The analytically calculated derivative is normalized so that its maximum value coincides with that of the transmitted signal. As a measure of the differentiation quality, we use the normalized root-mean-square deviation (NRMSD) of the absolute value of the transmitted beam from the analytically calculated derivative. It is evident from Fig. 3(b) that the differentiation quality is high, the NRMSD value in this case is only 1.2%. Note that in the NRMSD calculation, the shift of the transmitted beam was not taken into account (i.e. the minima of the transmitted beam and of the exact derivative were shifted to the same point). The maximum of the amplitude of the transmitted beam for this example equals 0.123.

By choosing the quality factor of the resonance, one can achieve the required tradeoff between the differentiation quality and the energy (amplitude) of the transmitted beam. For example, let us consider another resonant point $(l_{\text{zero},2},\theta_{\text{zero},2}) = (0.24\ \mu\text{m}, 50.14°)$ also marked with a black asterisk in Fig. 2(b), at which the transmission coefficient also vanishes. At this point, the resonant dip is narrower (i.e. the resonance has a higher quality factor). Therefore, the corresponding TF shown in Fig. 3(c) has a smaller linearity interval, but a larger amplitude. Thus, in this case, we should expect a decrease in the differentiation quality accompanied by an increase in the amplitude of the transmitted beam. The absolute value of the calculated profile of the transmitted beam corresponding to the point $(l_{\text{zero},2},\theta_{\text{zero},2})$ is shown in Fig. 3(d). In this case, the NRMSD value is indeed higher than in the previous example and amounts to 3.5%. At the same time, maximum amplitude of the transmitted signal is also higher and equals 0.21.

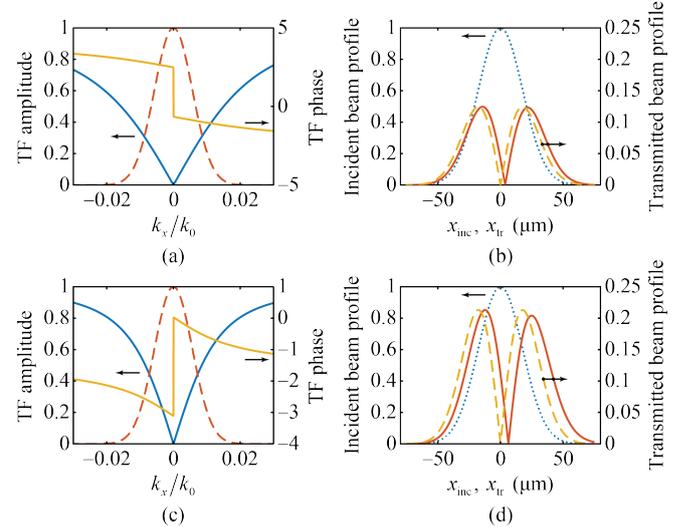

Fig. 3. Amplitudes (absolute values) and phases (arguments) of the transfer functions $H_T(k_{x,\text{inc}})$ calculated at the points $(l_{\text{zero},1},\theta_{\text{zero},1}) = (0.2\ \mu\text{m}, 48.62°)$ (a) and $(l_{\text{zero},2},\theta_{\text{zero},2}) = (0.24\ \mu\text{m}, 50.14°)$ (c). Dashed curves in (a) and (c) show the normalized spectrum of the incident Gaussian beam. (b), (d) Absolute values of the calculated profiles of the transmitted beams corresponding to the TFs shown in (a) and (c), respectively. Dotted curves show the incident Gaussian beam, dashed curves show the absolute value of the analytically calculated derivative of the Gaussian function.

## V. SPATIAL INTEGRATION OF A BEAM PROPAGATING IN THE WAVEGUIDE

The presence of high-Q resonances in the spectra in Fig. 2 makes it possible to use the considered structure as an optical spatial integrator operating in reflection [7], [10]. Indeed, in the vicinity of the resonance, the transmission coefficient can be approximated by the Fano lineshape [7]

$$R_{\text{TE}}(k_x) \approx r + \frac{b}{k_x - k_{x,\text{p}}}, \quad (7)$$

where $r$ is the non-resonant transmission coefficient, $b$ is the coefficient describing the coupling of the incident light to an eigenmode supported by the structure (by the rib waveguide), and $k_{x,\text{p}} = k'_{x,\text{p}} + ik''_{x,\text{p}}$ is the complex propagation constant of the eigenmode corresponding to the pole of the function $R_{\text{TE}}(k_x)$. Under the assumptions that $k_{x,0} = k'_{x,\text{p}}$, $R_{\text{TE}}(k_{x,0}) = 1$, and that the non-resonant reflection coefficient $r$ in (7) can be neglected, we can write the TF $H_R(k_{x,\text{inc}})$ of (3) in the following form:

$$H_R(k_{x,\text{inc}}) = \frac{\gamma}{k_{x,\text{inc}} - i\gamma}, \quad (8)$$

where $\gamma = k''_{x,\text{p}}/\cos\theta_0$. The TF in (8) is an approximation of

the TF of a perfect integrator with respect to the spatial variable $H_{\text{int}}(k_{x,\text{inc}}) = 1/(ik_{x,\text{inc}})$. As the Q-factor of the resonance increases (i.e. as the imaginary part of the mode propagation constant $k''_{x,\text{p}}$ decreases), the accuracy of the approximation also increases.

As an example, solid curves in Figs. 4(a) and 4(c) show the amplitudes of the TFs $H_R(k_{x,\text{inc}})$ calculated at two resonant points $(l_{\text{unity},1}, \theta_{\text{unity},1}) = (0.355\ \mu\text{m}, 53.28°)$ and $(l_{\text{unity},2}, \theta_{\text{unity},2}) = (0.36\ \mu\text{m}, 53.37°)$, at which the reflectance reaches unity. These points are marked with white asterisks in the inset of Fig. 2(a). The resonant approximations (8) are shown with dashed curves in Figs. 4(a) and 4(c) and are in a good agreement with the calculated TFs. The parameters of the approximations $\gamma = \gamma_1 = 0.0011\ \mu\text{m}^{-1}$ (point 1) and $\gamma = \gamma_2 = 0.0018\ \mu\text{m}^{-1}$ (point 2) were found by fitting (8) to the computed angular spectra.

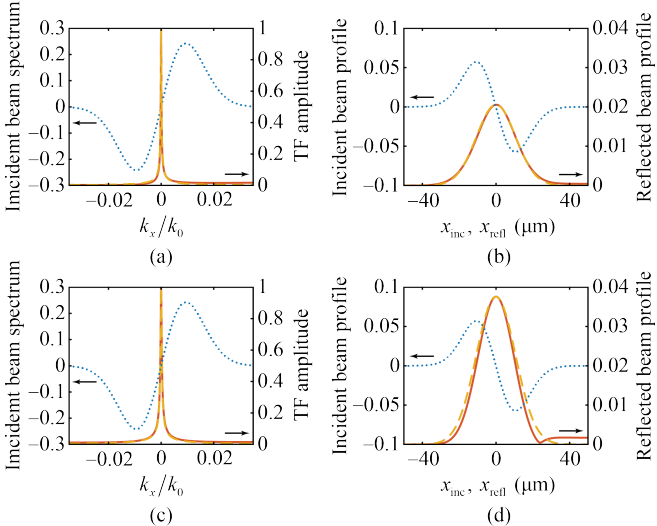

Fig. 4. Amplitudes (absolute values) of the transfer functions $H_R(k_{x,\text{inc}})$ calculated at the points $(l_{\text{unity},1}, \theta_{\text{unity},1}) = (0.355\ \mu\text{m}, 53.28°)$ (a) and $(l_{\text{unity},2}, \theta_{\text{unity},2}) = (0.36\ \mu\text{m}, 53.37°)$ (c). Dotted curves in (a) and (c) show the spectrum of the incident beam, dashed curves show the approximations of the TFs calculated using Eq. (8). (b), (d) Absolute values of the calculated profiles of the reflected beams corresponding to the TFs shown in (a) and (c), respectively. Dotted curves in (b) and (d) show the incident beam, dashed curves show the absolute values of the analytically calculated integral.

Let us now consider the integration of the beam propagating in the waveguide and having the profile $P_{\text{inc}}(x_{\text{inc}}) = (-2x_{\text{inc}}/\sigma^2)\exp(-x_{\text{inc}}^2/\sigma^2)$, which corresponds to the derivative of a Gaussian function $g(x_{\text{inc}}) = \exp(-x_{\text{inc}}^2/\sigma^2)$. The spatial spectrum of the beam $G(k_{x,\text{inc}}) \sim k_{x,\text{inc}}\exp(-k_{x,\text{inc}}^2\sigma^2/4)$ at $\sigma = 15\ \mu\text{m}$ is shown with a dashed curve in Figs. 4(a) and 4(c). In order to assess the integration quality, we calculated the profile of the reflected beam $P_{\text{refl}}(x_{\text{refl}})$ using (5). Figs. 4(b) and 4(d) show the incident beam (dotted curves), analytically calculated integral of the incident beam profile (dashed curves), and absolute values of the profiles of the reflected beams (solid curves) corresponding to the TFs of Figs. 4(a) and 4(c), respectively. The reflected beams are normalized by the maximum amplitude of the incident beam. The analytically calculated integral (the function $g(x_{\text{refl}})$) is shown scaled so that its value coincides with the amplitude of the reflected signal at $x_{\text{refl}} = 0$.

The resonance at the point $(l_{\text{unity},1}, \theta_{\text{unity},1})$ [Fig. 4(a)] has a higher Q-factor than the resonance at the point $(l_{\text{unity},2}, \theta_{\text{unity},2})$ [Fig. 4(c)]. Therefore, one can expect the quality of integration at the point $(l_{\text{unity},1}, \theta_{\text{unity},1})$ [Fig. 4(b)] to be higher than that at the point $(l_{\text{unity},2}, \theta_{\text{unity},2})$ [Fig. 4(d)]. Similarly to the previous section, we use NRMSD of the reflected beam from the analytically calculated integral as the measure of the quality of integration. For the considered examples, the NRMSD values amount to 1.2% [Fig. 4(b)] and 4.2% [Fig. 4(d)]. At the same time, the higher the integration quality (the higher the Q-factor), the lower the amplitude of the reflected signal. Maximum amplitudes of the reflected signal normalized by the maximum amplitude of the incident beam in the considered cases equal 0.02 [Fig. 4(b)] and 0.038 [Fig. 4(d)]. Thus, similarly to the differentiation case, it is possible to achieve the required tradeoff between the quality of integration and the energy (amplitude) of the reflected beam by choosing the Q-factor of the resonance.

## VI. Conclusion

In the present work, we demonstrated that in the case of oblique incidence of TE-polarized modes of a slab waveguide on a dielectric ridge located on the waveguide surface, resonant changes in the reflectance and the transmittance occur. Using the effective index method, we explained these resonant effects by the excitation of a cross-polarized mode of the ridge.

The discovered resonances enable optical implementation of the operations of spatial differentiation and integration of the profile of an optical beam propagating in the waveguide. The computation of the derivative is performed in transmission, whereas the computation of the integral is performed in reflection. Rigorous numerical simulation results confirm the possibility of high-quality spatial differentiation and integration.

The obtained results may find application in the design of on-chip systems for all-optical analog computing.

The authors also believe that the presented planar structure can be used as an integrated optical spatial (angular) or spectral filter. Moreover, the existence of the bound states in the continuum in the structure also makes it promising for lasing, sensing and enhancing non-linear light-matter interactions. A detailed investigation of the BIC formation mechanism in this structure will be the subject of a future

**Evgeni A. Bezus** graduated with honors from Samara State Aerospace University (presently, Samara National Research University (Samara University)), Samara, Russia, in 2009, majoring in Applied Mathematics and Computer Science. He received his Candidate of Science in Physics and Mathematics degree from Samara State Aerospace University in 2012. Currently, he is a researcher at the Diffractive Optics laboratory of the Image Processing Systems Institute (IPSI RAS – Branch of the FSRC "Crystallography and Photonics RAS"), Samara, Russia, and an associate professor at Technical Cybernetics department of Samara University. His research interests include nanophotonics, plasmonics, and electromagnetic diffraction theory.

**Leonid L. Doskolovich** graduated with honors from the S.P. Korolyov Kuybyshev Aviation Institute (presently, Samara University), Samara, Russia, in 1989, majoring in Applied Mathematics. He received his Doctor of Science in Physics and Mathematics (2001) degree from Samara State Aerospace University. Currently, he is the Head of Diffractive Optics Laboratory of the Image Processing Systems Institute, Samara, Russia, and a professor at Technical Cybernetics department of Samara University. His research interests include diffractive optics, nanophotonics, and plasmonics.

**Dmitry A. Bykov** graduated with honors from the Samara State Aerospace University, Samara, Russia, in 2009, majoring in Applied Mathematics and Computer Science. He received his Doctor of Science in Physics and Mathematics degree from Samara University in 2017. Currently he is a senior researcher at the Diffractive Optics Laboratory of the Image Processing Systems Institute and an associate professor at Technical Cybernetics department of Samara University. His research interests include nanophotonics, magnetooptics of nanostructured materials, plasmonics and grating theory.

**Victor A. Soifer** graduated from Kuybyshev Aviation Institute (presently, Samara University) in 1968. He received his Doctor of Science in Engineering degree from Saint Petersburg State Electrotechnical University "LETI" in 1979. He is an Academician of the Russian Academy of Sciences (RAS) since 2016. Currently, he is the President of Samara University and a chief researcher at the IPSI RAS – Branch of the FSRC "Crystallography and Photonics" RAS. His research interests include diffractive optics and nanophotonics. V. A. Soifer is a member of SPIE and IAPR.